\documentclass[sigconf,nonacm]{acmart}
\usepackage{ifthen}      
\usepackage{moreverb,url}
\usepackage{algorithm}
\usepackage{algorithmic}
\usepackage{subcaption}
\usepackage{amsmath}
\usepackage{amssymb}
\usepackage{dsfont}
\usepackage{mathrsfs}
\usepackage{color}
\usepackage{newfloat}
\usepackage{listings}

\usepackage{enumitem}
\usepackage{booktabs}
\usepackage{xcolor}
\usepackage{graphicx}
\usepackage{tabularx}
\usepackage[inkscapelatex=false]{svg}
\usepackage{lipsum}

\newboolean{isanonymized}
\setboolean{isanonymized}{true} 

\newcommand{\anonymize}[2]{%
    \ifthenelse{\boolean{isanonymized}}{#1}{#2}%
}

\providecommand{\anon}[1]{}
\renewcommand{\anon}[1]{%
    \anonymize{[Anonymized for peer review]}{#1}%
}

\copyrightyear{2027}
\acmYear{2027}
\setcopyright{acmlicensed}
\acmConference[WWW '27]{Proceedings of the ACM Web Conference 2027}{2027}{TBD}
\acmBooktitle{Proceedings of the ACM Web Conference 2027 (WWW '27)}
\acmDOI{10.1145/nnnnnnn.nnnnnnn}
\acmISBN{978-x-xxxx-xxxx-x/YY/MM}

\newcommand{\Unq}{\mathrm{Unq}}
\newcommand{\Red}{\mathrm{Red}}
\newcommand{\Syn}{\mathrm{Syn}}

\newcommand{\CoI}{\mathrm{CoI}}
\newcommand{\IP}{\mathrm{IP}}
\newcommand{\EI}{\mathrm{EI}}

\newcommand{\own}{\mathrm{own}}
\newcommand{\fld}{\mathrm{field}}
\newcommand{\bias}{b}
\newcommand{\owner}{o}

\newcommand{\ROneGdeltPooledSameOwnerRatio}{20.4}  
\newcommand{\ROneCcPooledSameOwnerRatio}{2.2}  

\newcommand{\RTwoEIOwnerSD}{0.72}  
\newcommand{\RTwoEIAffSD}{0.15}  
\newcommand{\RTwoEIAPSlope}{-0.063}  
\newcommand{\RTwoEIVSOwner}{0.258}  
\newcommand{\RTwoEIOwners}{61}  
\newcommand{\RTwoEILowestOwnerBLUP}{-1.53}  
\newcommand{\RTwoEIHOwnerSD}{0.76}  
\newcommand{\RTwoEIHAPSlope}{-0.210}  
\newcommand{\RTwoIPOwnerSD}{0.68}  
\newcommand{\RTwoIPAffSD}{0.14}  
\newcommand{\RTwoDwOwnerSD}{0.92}  
\newcommand{\RTwoDwAffSD}{0.15}  
\newcommand{\RTwoDwVSOwner}{0.436}  
\newcommand{\RTwoVolEIOwnerSD}{0.76}  
\newcommand{\RTwoVolEIHOwnerSD}{0.75}  
\newcommand{\RTwoVolIPOwnerSD}{0.59}  
\newcommand{\RThreeAlphaTVSix}{0.74}  
\newcommand{\RThreeAlphaPrintTwelve}{0.83}  
\newcommand{\RThreeBetaFailedTwelve}{-0.07}  
\newcommand{\RThreeBetaFailedDeals}{11}  
\newcommand{\RThreeHzAcqNTreated}{212}  
\newcommand{\RThreeHzAcqNControl}{297}  
\newcommand{\RThreeKmAcqEITreatSix}{54}  
\newcommand{\RThreeKmAcqEICtrlSix}{26}  
\newcommand{\RFourAirExcessOwnerWithinNet}{-0.022}  
\newcommand{\RFourAirExcessNetAmongDiffOwner}{+0.257}  
\newcommand{\RFourWebExcessOwnerWithinNet}{+0.00788}  
\newcommand{\RFiveDecisionRawDiff}{+0.0067}  
\newcommand{\RFiveDecisionRawP}{0.00071}  
\newcommand{\RFiveNBlockedDeals}{6}  
\newcommand{\RFiveNClearedDeals}{47}  
\newcommand{\RFiveApprovedSumSD}{+2.0}  
\newcommand{\ROnePlacePPooledSameOwnerRatio}{40.2}  
\newcommand{\ROnePlacePLeadK}{10}  

\newcommand{\RTwoArticlesOwnerSD}{0.61}  
\newcommand{\RTwoBylinesOwnerSD}{0.73}  
\newcommand{\RTwoTrafficOwnerSD}{0.88}  
\newcommand{\RThreeKmLayoffAuthorsTreatSix}{44}  
\newcommand{\RThreeKmLayoffAuthorsCtrlSix}{33}  
\newcommand{\RThreeHzEcAcqEISix}{2.12}  
\newcommand{\RThreeHzEcAcqEISixLo}{1.47}  
\newcommand{\RThreeHzEcAcqEISixHi}{3.07}  
\newcommand{\RThreeHzEcAcqEITwentyFour}{1.55}  
\newcommand{\RThreeHzEcAcqEITwentyFourLo}{1.02}  
\newcommand{\RThreeHzEcAcqEITwentyFourHi}{2.34}  
\newcommand{\RThreeHzEcAcqIPSix}{0.61}  
\newcommand{\RThreeHzEcAcqIPSixLo}{0.23}  
\newcommand{\RThreeHzEcAcqIPSixHi}{1.61}  
\newcommand{\RThreeHzEcAcqAuthorsTwentyFour}{0.67}  
\newcommand{\RThreeHzEcAcqAuthorsTwentyFourLo}{0.53}  
\newcommand{\RThreeHzEcAcqAuthorsTwentyFourHi}{0.85}  
\newcommand{\RThreeHzPlaceLayoffIPThree}{3.09}  
\newcommand{\RThreeHzPlaceLayoffIPThreeLo}{1.22}  
\newcommand{\RThreeHzPlaceLayoffIPThreeHi}{7.84}  
\newcommand{\RThreeHzPlaceAcqEINTreated}{28}  
\newcommand{\RThreeHzPlaceAcqIPNTreated}{12}  
\newcommand{\RThreeHzPlaceLayoffIPNTreated}{46}  
\newcommand{\RThreeHzCcAcqEITwoGraphs}{5.81}  
\newcommand{\RThreeHzCcAcqEITwoGraphsLo}{2.16}  
\newcommand{\RThreeHzCcAcqEITwoGraphsHi}{15.64}  
\newcommand{\RThreeHzCcAcqEIEightGraphs}{3.56}  
\newcommand{\RThreeHzCcAcqEIEightGraphsLo}{1.67}  
\newcommand{\RThreeHzCcAcqEIEightGraphsHi}{7.58}  
\newcommand{\RThreeHzCcLayoffIPFourGraphs}{0.45}  
\newcommand{\RThreeHzCcLayoffIPFourGraphsLo}{0.21}  
\newcommand{\RThreeHzCcLayoffIPFourGraphsHi}{0.95}  
\newcommand{\RThreeHzCcAcqEINTreated}{88}  

\newcommand{\RTwoCcEIOwnerSD}{1.05}  
\newcommand{\RTwoCcEIAffSD}{0.06}  
\newcommand{\RTwoCcEIOutlets}{595}  
\newcommand{\RTwoPlaceEIOwnerSD}{0.69}  
\newcommand{\RTwoPlaceEIAffSD}{0.16}  
\newcommand{\RTwoPlaceEIOutlets}{656}  

\newcommand{\changedtext}[1]{#1}
\newcommand{\vone}[1]{#1}

\colorlet{BLUE}{blue}  
\newenvironment{changed}{}{}

\title[Organizational Influence on Digital News]{Measuring Organizational Influence on Digital News: An Information-Theoretic Framework for Media Pluralism and Editorial Independence}

\author{Peter Carragher}
\affiliation{%
  \institution{Carnegie Mellon University}
  \city{Pittsburgh}
  \state{Pennsylvania}
  \country{USA}}
\email{pcarragh@andrew.cmu.edu}

\author{Kathleen M. Carley}
\affiliation{%
  \institution{Carnegie Mellon University}
  \city{Pittsburgh}
  \state{Pennsylvania}
  \country{USA}}
\email{carley@andrew.cmu.edu}

\begin{document}

\begin{abstract}
News consolidation is accelerating, yet regulators have no measurement frameworks for determining whether owners shape editorial decisions. Existing evidence comes from single cases, where pluralism and independence are measured separately, if at all. We ask whether ownership impacts editorial decisions across the digital news ecosystem, and if so, how acquisitions affect this. We derive editorial independence and internal pluralism using partial information decomposition, which separates news coverage overlap with sister outlets from coverage overlap with all outlets. Using public datasets, we apply our measures monthly from 2018 to 2026 to 1,364 US news websites on three editorial decisions: which events they cover, which sources they link to, and what they put on their front pages. We find that co-owned outlets share up to 40 times more information about these decisions than expected by chance. Ownership accounts for far more variation in independence than network affiliation does and, upon being acquired, outlets lose independence sooner than matched controls, quickly converging on their new owner. In broadcast news, by comparison, we find that it is network affiliation, not ownership, that accounts for more content overlap. Finally, we discuss policy implications and suggest that media regulators have blocked large deals rather than editorially detrimental ones. We hope that this work can serve as a framework for comparative media analyses going forward.

\end{abstract}

\ccsdesc[500]{Information systems~Web mining}
\ccsdesc[300]{Social and professional topics~Governmental regulations}
\ccsdesc[500]{Mathematics of computing~Information theory}

\keywords{News media, media ownership, editorial independence, media pluralism, information theory, GDELT, Common Crawl}

\maketitle

\section{Introduction}

Paramount Skydance's \$110 billion acquisition of Warner Bros.\ Discovery, approved by the DOJ and the FCC, places CBS
News and CNN under one owner. Together, their outlets now reach 97--99\% of US households. 
Given concerns from media critics, a five-journalist editorial board to protect the two newsrooms' independence is being established.

Yet it is unclear how such a board would know whether independence has been maintained, or what would erode it.
While research on organizational influences over editorial decisions is well established, and there is general agreement that such influence exists, current research falls short of measuring it.
Existing studies rely on case studies and adopt heterogeneous methods for analysis \citep{theine2025does}, treating pluralism and independence as unrelated measures.
What's more, regulators seem to approve acquisitions based on audience reach and local market overlap, disregarding editorial decisions.
No measure tells a regulator before a deal, or an editorial board after it, the degree to which editorial decisions may change. 
In terms of transparency for the information ecosystem, no comparative measures exist to inform the public about which news outlets have the least editorial independence.

However, editorial decisions leave traces that can be measured at Web scale, such as which events an outlet covers (GDELT), which sources it cites (Common Crawl web graphs), and which stories it puts on its front page (GDELT Global
Frontpage Graph). Because editorial decisions are documented at Web scale, they can be used to audit owner influence,
which is a question of information integrity and media governance.
We ask four questions. Does ownership organize outlets' editorial decisions? If so, how much, relative to network
affiliation? Do organizational changes such as acquisitions and layoffs impact editorial independence? And do regulators block the deals that most reduce editorial independence? 
We define pluralism and independence from
the same primitives: an outlet's monthly editorial decisions, and the information that its sibling outlets, and the rest of
the ecosystem, convey about them. A partial information decomposition separates what an outlet shares specifically with
its siblings from what it shares with any other outlet; with political bias as the grouping, the measure is internal
pluralism, and with ownership, editorial independence. 
We demonstrate that pre-existing measurement methods cannot account for this and instead mistake the common news agenda, such as content from wire services such as the Associated Press (AP News), as evidence of low editorial independence between sibling stations.
We then apply our measures to Web-scale datasets covering 1,364 US news websites, monthly from 2018 to 2026.
A generalized additive mixed model sizes the owner effect against network affiliation, and matched survival designs
follow outlets after acquisitions and layoffs.

Compared to the shared information between two random outlets, co-owned outlets share $\sim 20 \times$ more information for event selection,
$\sim 2 \times$ more for (unweighted) source selection, and 
$\sim 40 \times$ more for front-page placement. 
Meanwhile, outlets that share only a network affiliation have reduced, though still significant, overlap.
Owners account for far more variation in independence than affiliate networks (owner SD \RTwoEIOwnerSD{}, affiliate SD \RTwoEIAffSD).
This effect depends on media type, however, and we find that it reverses when considering content from television broadcasts. 
On air, stations' shared coverage follows their network affiliation; online it follows their owners.

Acquired outlets lose independence sooner than matched controls (hazard ratio \RThreeHzEcAcqEISix{} at six months), 
quickly converging on their new owner. 
Projecting these trends over past deals, we suggest that regulators have blocked large deals rather than ones that stand to reduce editorial independence. 
In this way, our measures can track acquired newsrooms over time, giving editorial boards such as the
one for CBS News and CNN a benchmark. We contribute a unified framework for editorial independence and
pluralism on three editorial decisions, the findings derived from them, and an ownership panel of 1,364 US news websites.
\section{Related Work}
\label{sec:related}

Studies of organizational influence on editorial decisions broadly agree that such influence exists. 
Of the 56 empirical studies reviewed by \citet{theine2025does}, 33 report strong ownership influence, 13 weak influence, seven no systematic
relationship, and three are unclear. However, because the studies are methodologically heterogeneous and mostly examine a few outlets, a comparative analysis gap remains.
\paragraph{Pluralism.}
Comparative media research distinguishes \emph{external} pluralism, a system of outlets that each carry a different
political tendency, from \emph{internal} pluralism, balance within each outlet
\citep{hallin2004comparing,mcquail1993media}. Content studies measure these with diversity indices such as Shannon entropy over various categories
of content like topics, actors, and viewpoints \citep{masini2018measuring,loecherbach2020unified}.

Whether an index measures internal or external pluralism depends on whether it is applied to a single outlet's
distribution or to the whole market. We apply this decomposition to editorial decisions. At time $t$ and for a given type of editorial decision, with $X_i(e)$ indicating that outlet $i \in O_t$ covered item
$e \in E_t$ (an event, or a domain it cites), let $p_i(e) = X_i(e) / \sum_{e'} X_i(e')$ be outlet $i$'s coverage distribution,
$w_i = \sum_{e} X_i(e) / \sum_{j,e} X_j(e)$ its share of all coverage, and $\bar p = \sum_i w_i p_i$ the ecosystem-level coverage distribution. 
Total diversity decomposes into the diversity of each outlet's coverage distribution separately and the differences between those distributions
\citep{jost2007partitioning}:
\begin{equation}
  H(\bar p) \;=\; \underbrace{\textstyle\sum_{i \in O_t} w_i\, H(p_i)}_{\text{within outlets}}
  \;+\; \underbrace{\textstyle\sum_{i \in O_t} w_i\, D_{\mathrm{KL}}(p_i \,\|\, \bar p)}_{\text{between outlets}} .
  \label{eq:partition}
\end{equation}

A system is externally plural when total diversity is high and mostly supplied by the between-outlet term,
and internally plural when it is mostly supplied by the within-outlet term. The between-outlet term is the mutual
information between two outlets in terms of the items they cover, so external pluralism is high when outlets select different items. 
However, this formulation alone does not explain why some outlets select alike and others do not. 
To answer this, our measure introduces a further decomposition where we split the
shared coverage information into the part specific to an outlet's own group (its owner or political bloc) and the part
it shares with the rest of the media ecosystem (Section~\ref{sec:methods}).


\paragraph{Ownership and editorial independence.}
Policy monitors use institutional safeguards \citep{mpm2023} and
survey studies from expert ratings \citep{popescu2011emss,hanretty2014media} to score editorial independence at the country level.
Studies that measure an owner's influence over its outlets' output do so in two ways. The first is the owner's share
of the variance in an outlet trait, as with the owner random effect on newspaper slant of \citet{gentzkow2010drives}.
The second is the change in coverage after an acquisition, relative to
comparison outlets \citep{archer2018changing,martin2019local,mastrorocco2025who}. The review of \citet{theine2025does} names online news and
cross-national comparison as gaps in this literature, which this work addresses. We estimate the owner's variance share for each editorial decision with our central model (Section~\ref{sec:model-gamm}).

The closest outlet-level measure of independence is content recycling, the share of an outlet's article titles
that also appear in co-owned outlets \citep{hendrickx2021why}. With $X_i(e)$ as in \eqref{eq:X}, let $Z^{\own}_i(e)$ indicate that some other outlet in $i$'s owner group did (Section~\ref{sec:groupings}). Content recycling is then $P(Z^{\own}_i > 0 \mid X_i = 1)$. 

\begin{figure}
    \centering
    \includegraphics[width=0.45\linewidth]{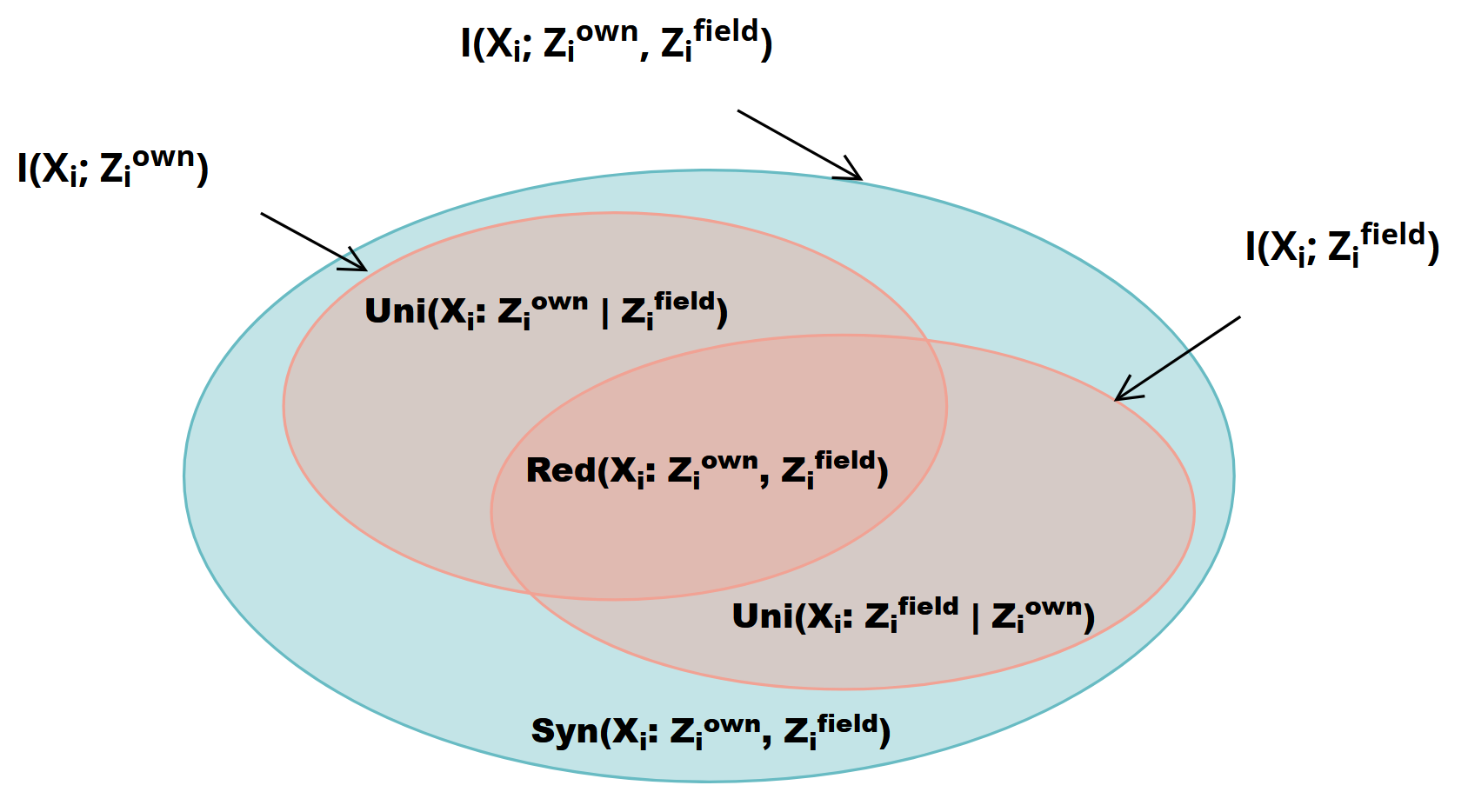}
    \caption{The PID of the information that outlet $i$'s co-owned siblings have about $i$'s editorial decisions. $\Unq_i^{\own}$ is unique information, what only $i$'s group predicts about its coverage, $\Red_i$ is redundancy, what group and field both predict, and $\Syn_i$ is synergy, what only their combination predicts.}
    \label{fig:pid_figure}
\end{figure}

Crucially, this measure can present ecosystem-level redundancy as co-ownership overlap. 
In contrast, our partial information decomposition (PID) \citep{williams2010nonnegative} separates the two: the information about an outlet's coverage that is conveyed by its owner is split into what the owner uniquely provides ($\Unq^{\own}_i$) and what the rest of the ecosystem also provides
(redundancy $\Red_i$; \autoref{fig:pid_figure}; Section~\ref{sec:methods}):
\begin{equation}
  I\bigl(X_i ; Z^{\own}_i\bigr) \;=\; \Unq^{\own}_i + \Red_i .
  \label{eq:recycling_pid}
\end{equation}



We apply the same construction with political bias in place of ownership, so internal
pluralism and editorial independence are computed from a unified framework. 
While several web-scale studies use political bias to describe news ecosystems from hyperlink and co-coverage networks
\citep{rappaz_dynamic_2019,misinformation_discovery,www_bias_detection}, we know of no work that measures pluralism
and independence together.
\section{Measures}
\label{sec:methods}

We define outlet measures as ratios of the partial information decomposition (PID) atoms of the outlet's
coverage, with pairwise quantities derived with respect to the partial entropy decomposition (PED). 
Each measure is defined for both the political bias and parent owner outlet groupings. For a mapping between the two sets of measures, see Appendices~\ref{app:methods-details} and~\ref{app:mi}.

\subsection{Coverage variables}
\label{sec:variables}
\label{sec:groupings}
An outlet is active in window $t$ (a GDELT month, a Common Crawl graph, or a Sunday front page) if it covers at least 50 items in it; $O_t$ is
the set of active outlets. Items are GDELT events, cited domains or front-page events, each representing a different editorial decision
(Section~\ref{sec:decisions}). For an item $e$ drawn uniformly at random,
\begin{equation}
  X_i(e) \;=\; \mathbf{1}\bigl[\text{outlet } i \text{ covered item } e\bigr], \qquad i \in O_t,
  \label{eq:X}
\end{equation}
where covering means publishing an article about the event, or linking to the domain. 
To reduce noise from overly granular event coding in GDELT, we filter out
the items covered by fewer than $m = 3$ active outlets:
\begin{equation}
  E_t^{(m)} \;=\; \Bigl\{ e : \textstyle\sum_{i \in O_t} X_i(e) \ge m \Bigr\}.
  \label{eq:filter}
\end{equation}
Because the filter itself induces dependence between outlets, we test the measures against a null model where they are recomputed after conducting degree-preserving randomization on the underlying data, which preserves each outlet’s number of items and each item’s number of covering outlets (Appendix~\ref{app:methods-details}).

A grouping $C \in \{\bias, \owner\}$ assigns outlet $i$ to a group $G_{c(i)}$: its bias category, or its owner in month
$t$. Two \emph{sources} describe each item from $i$'s point of view:
\begin{equation}
  Z^{\own}_{i}(e) = \operatorname{bin}\Bigl( \textstyle\sum_{j \in G_{c(i)} \setminus i} X_j(e) \Bigr), \quad
  Z^{\fld}_{i}(e) = \operatorname{bin}\Bigl( \textstyle\sum_{j \notin G_{c(i)}} X_j(e) \Bigr),
  \label{eq:sources}
\end{equation}
how many other outlets of $i$'s group covered $e$, and how many outlets outside the group did, binned on a log scale. Binning is used to restrict $Z$ to a six-state variable, so that estimation of the PID’s joint table is tractable.
Under $C = \bias$, $i$’s co-owned outlets in the same bias category are dropped from both sources, so bias and ownership are not conflated.

\subsection{Outlet measures}
\label{sec:decomp}
\label{sec:family1}
The information the two sources carry about $i$'s coverage is split by PID \cite{williams2010nonnegative,bertschinger2014quantifying} into four non-negative atoms (\autoref{fig:pid_figure}):
\begin{equation}
  I_i \;=\; I\bigl(X_i ; Z^{\own}_i, Z^{\fld}_i\bigr) \;=\; \Unq_i^{\own} + \Unq_i^{\fld} + \Red_i + \Syn_i .
  \label{eq:pid}
\end{equation}
Internal pluralism ($IP$) and editorial independence ($EI$) are the shares of the explained information that is \emph{not} specific to
the group:
\begin{equation}
  \IP_i = 1 - \frac{\Unq_i^{\own}}{I_i}\ (C = \bias), \qquad
  \EI_i = 1 - \frac{\Unq_i^{\own}}{I_i}\ (C = \owner).
  \label{eq:ip}
\end{equation}
Hence, an outlet that covers exclusively what every outlet covers still has a high $EI$.
The characteristic that distinguishes $EI$ from content recycling (Section~\ref{sec:related}) is that shared-agenda coverage of this nature is
assigned to $\Red_i$ instead of $\Unq_i^{\own}$. 
To demonstrate this distinction, we also report the share of $i$'s coverage
entropy that its full set of co-owned outlets leaves unexplained: $\EI^H_i = H(X_i \mid X_{G_{c(i)} \setminus i}) / H(X_i)$ ($C = \owner$). As in content recycling (Section~\ref{sec:related}), $EI^H_i$ counts shared-agenda coverage as dependence.

\subsection{Pairwise measures}
For two outlets, the PED of their joint coverage has a shared atom $h_{\{i\}\{j\}}$ and a synergistic atom
$h_{\{ij\}}$, and their mutual information is the difference, $I(X_i;X_j) = h_{\{i\}\{j\}} - h_{\{ij\}}$ \citep{ince2017ped}.
We compare outlet pairs by the normalized mutual information
\begin{equation}
  \mathrm{NMI}_{ij} \;=\; \frac{I(X_i;X_j)}{\min\{H(X_i), H(X_j)\}} ,
  \label{eq:nmi}
\end{equation}
the share of the smaller coverage entropy that the two outlets share. Weighting the ordinal distance between bias
categories, $d(c,c') = |c - c'|/4$, by this coupling gives $i$'s cross-bias breadth,
$D_{w,i} = \sum_{j \neq i} d(c(i),c(j))\, I(X_i;X_j) / \sum_{j \neq i} I(X_i;X_j)$.
Unlike $IP_i$, $D_{w,i}$ uses the ordering of bias categories, where coupling with a distant category counts more than coupling with a neighboring one. 
Building this ordering into the PID would require a group source, $Z$, with one state per category. 
While not intractable, this would make PID estimates unreliable at our sample sizes (Appendix~\ref{sec:estimation}).
We therefore keep the PID measures nominal and only use ordinal variables for political bias in $D_{w,i}$.


\section{Data}
\label{sec:data}

\paragraph{Event selection.} Event selection comes from the Mentions table of the Global Database of
Events, Language, and Tone (GDELT) \citep{leetaru2013gdelt}, which records the source domains that published articles
about each news event that GDELT extracts. We aggregate these events by calendar month. 
During periods where chain-owned sites blocked GDELT's crawlers, we mask the affected outlet-months.

\paragraph{Source selection.} Source selection comes from the Common Crawl
domain-level web graphs \citep{commoncrawl2024}, which record links between domains. We parse each of the 55 web graphs published
since February 2018 as a separate snapshot. 

\paragraph{Front-page placement.} Front-page placement comes from the GDELT Global Frontpage
Graph (GFG) \citep{gdelt2018gfg}, which snapshots the links on about 50{,}000 news homepages every hour and records each
link's position on the page. We use one Sunday a month, starting in 2018. We note that, as of October 2025, GDELT appears to have stopped actively maintaining the GFG. Measurements made on this data after that point are therefore unreliable.

\paragraph{Outlets and other outcomes.} Outlets are the US news domains of our ownership panel, which
labels each domain with its parent owner month by month and with its network affiliations
(Appendix~\ref{app:ownership}); political-bias labels in five ordinal categories come from Media Bias/Fact Check
\citep{zandt2024mbfc}. We also measure productivity (GDELT article and
byline-author counts) and engagement (Ahrefs organic web traffic estimates \citep{ahrefs}). 
Byline authors are the distinct names in the author field GDELT extracts from each article, split and normalized (lower case, letters
only) and counted per outlet-month. Wire dependence is the share of an outlet's bylined articles in the month that
credit the Associated Press, by name, as ``AP'' or as an AP desk.

\begin{figure*}
    \centering
    \includegraphics[width=\textwidth]{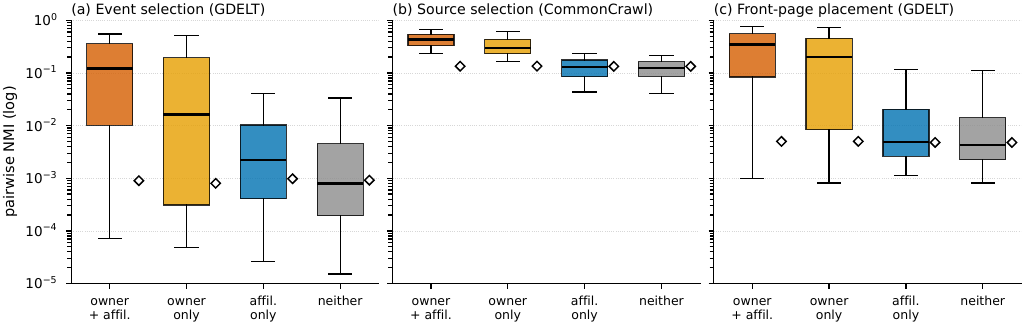}
    \caption{Shared editorial decisions by co-organizational type: shared owner and affiliation, shared owner only, shared affiliation only, and no relation. Each plot is the
    distribution of pairwise NMI \eqref{eq:nmi} between two outlets' monthly decisions. 
    Diamonds show the median of the owner- and affiliation-randomized null.} 
    
    \label{fig:ownership_timeseries_correlations}
    \label{fig:r1_similarity_by_coorg}
\end{figure*}

\paragraph{Organizational events.} Treatment groups for our survival analysis are made up of outlets from organizations that experienced layoffs of at least 100 jobs.
This data is collected from public trackers of newsroom cuts \vone{\citep{tobitt2026jobcuts}}. For US news media acquisitions and mergers, ranging from 2018 to mid-2023, we use pre-existing datasets \citep{martin2026media,LocalNews2025} (see Appendix~\ref{app:data-details} for more details). Failed deals (\RThreeBetaFailedDeals{} blocked or abandoned, 2017--2026) are compiled by hand from FCC orders, DOJ announcements, SEC filings and trade press.

\section{Model}
\label{sec:model}

\label{sec:model-gamm}

\begin{changed}
We fit one generalized additive mixed model per outcome $y_{it}$ (outlet $i$, month $t$) with \texttt{mgcv}
\citep{wood2017gam}:
\begin{equation}
\begin{aligned}
  y_{it} = {}& \beta_0 + \beta_1\,\mathrm{ap}_{it} + \beta_2\,\mathrm{ap}_{it}\,\mathrm{share}_{it} \\
  & + f(t) + u_{o(i,t)} + a_{n(i)} + g_i(t) + \varepsilon_{it},
\end{aligned}
  \label{eq:gamm}
\end{equation}
where $\mathrm{share}_{it}$ is the outlet’s AP byline share (defined when $\mathrm{ap}_{it}=1$, i.e., the outlet published bylined articles that month),
$f(t)$ a smooth of calendar time, $u_{o(i,t)}$ and $a_{n(i)}$ random intercepts for the owner of $i$ in month $t$
and its network affiliate, and $g_i(t)$ a per-outlet smooth of time.
Because each outlet has its own smooth, the owner effect measures what co-owned outlets have in common rather than differences between individual outlets.
Where residuals remain correlated from month to month, we refit with first-order autoregressive errors within outlets. 
Finally, the ownership effect is the owner random-effect SD in outcome-SD units, where zero variance indicates no owner
influence \citep{gentzkow2010drives}.
\end{changed}

\paragraph{Outcome measures.}
\label{sec:decisions}
Outcomes $y_{it}$ for outlet $i$ are $IP_i$ and $EI_i$ on each of three types of editorial decisions. We define the
items $e$ and $X_i(e)$ of \eqref{eq:X} as the GDELT events $i$ covered (event selection, monthly), domains $i$ links to in a
Common Crawl graph (source selection), and GDELT events linked from $i$'s homepage on one Sunday a month (front-page
placement).
\section{Results}

First, we establish the existence of organizational influence over three editorial decisions. Then, we turn to quantifying the level of influence using the GAMM \eqref{eq:gamm}, investigating the impact of organizational changes such as acquisitions and layoffs, and comparing digital with broadcast news. Finally, we explore how platform policies impact digital news outlets with a longitudinal analysis of online engagement data.

\subsection{Ownership and editorial decisions}
\label{sec:ownership}

Co-owned outlets make the same editorial decisions far more often than other outlets
(\autoref{fig:r1_similarity_by_coorg}). Same-owner pairs share \ROneGdeltPooledSameOwnerRatio{} times the information of
the shuffled-owner null in event selection, \ROneCcPooledSameOwnerRatio{} times in source selection and
\ROnePlacePPooledSameOwnerRatio{} times in front-page placement. In contrast, pairs that share only a network affiliation sit just
above the label-shuffled null, and share a significant proportion of their event-selection information only (\autoref{fig:r1_similarity_by_coorg}a).

We also see that median shared information between pairs of outlets that are co-owned and affiliated is relatively small for event selection (12\%).
In other words, 12\% of the information in the smaller outlet in these pairs is redundant given the other outlet.
In contrast, source selection ($\sim$43\%) and front-page placement ($\sim$35\%) have more shared information between co-owned and affiliated outlets.
For source selection, higher rates of shared information are a result of the higher density and connectivity of the web graph in comparison with the co-event coverage graph. Credible news outlets eventually link to one another at least once, so linking to the same
sources is common. This is evident in \autoref{fig:r1_similarity_by_coorg}b where even random outlets share 13\% of their information. We note that the Common Crawl web graph is binarized, so this analysis does not factor in link weights. In addition, NMI scores over GDELT events increase when computed between outlet groups of increasing size, as event coverage overlaps increase (see \autoref{fig:nmi_vs_group_size_gdelt_events} in the appendix).
For front-page placement, the higher rates of shared information likely result from a tradeoff between the limited space on the front page and the need for a news outlet to have major news stories (e.g., the shared agenda) represented there.

Across all three editorial decisions, a shared affiliation on its own implies a lack of shared information between outlets. We note, however, that this is primarily the case for digital news, and that media type plays an important role in moderating the role of network affiliations in editorial decision making. We return to this in Section~\ref{sec:validation}.

%


\subsection{Modeling organizational effects}

\begin{figure}
    \centering
    \includegraphics[width=\linewidth]{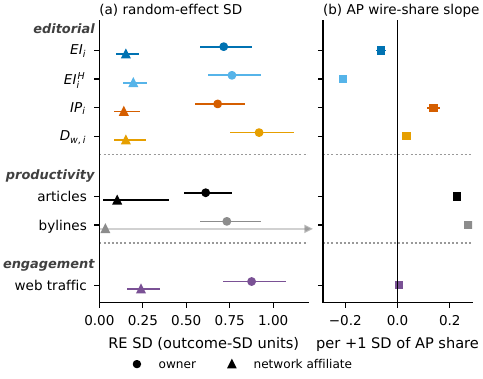}
    \caption{(a) Size of the owner and affiliate effects in the GAMM, by
    random-effect SDs in outcome-SD units, with 95\% intervals; (b) the AP wire-share slope per +1 SD of share.
    \changedtext{Editorial outcomes, productivity (articles, byline authors; log) and engagement (web traffic;
    asinh).}
    }
    \label{fig:gamm_coefficients_for_ownership}
\end{figure}

We apply our GAMM \eqref{eq:gamm} to size organizational effects (\autoref{fig:gamm_coefficients_for_ownership}). For every editorial
measure the owner SD is several times the affiliate SD ($EI_i$: \RTwoEIOwnerSD{} vs \RTwoEIAffSD; $IP_i$:
\RTwoIPOwnerSD{} vs \RTwoIPAffSD; $D_{w,i}$: \RTwoDwOwnerSD{} vs \RTwoDwAffSD), and co-ownership accounts for
\RTwoEIVSOwner{} to \RTwoDwVSOwner{} of the total variance. Here, SD is the standard deviation of the owner (or affiliate) random intercepts, in units of the outcome’s SD, so owners differ in their typical level of $EI_i$ by \RTwoEIOwnerSD{} $EI_i$ SDs.
Importantly, ownership influences productivity and traffic as much as editorial decisions. The owner SD is
\RTwoArticlesOwnerSD{} for article counts, \RTwoBylinesOwnerSD{} for byline authors and \RTwoTrafficOwnerSD{} for web traffic. 
Because productivity and engagement distributions are long-tailed, article and author counts enter as $\log(1+y)$ and web traffic as $\operatorname{arcsinh}(y)$.

AP wire sharing separates our measure from content recycling (Section~\ref{sec:related}): a higher AP share lowers $EI^H_i$,
which counts shared-agenda coverage as dependence, about three times as much as $EI_i$ (\RTwoEIHAPSlope{} vs
\RTwoEIAPSlope{}, \autoref{fig:gamm_coefficients_for_ownership}b). AP wire sharing is also associated with a modest increase in $IP_i$.

The larger influence of ownership relative to affiliation holds on the other two editorial decisions (\autoref{fig:gamm_other_decisions}, Appendix~\ref{app:other-decisions}): the owner SD of $EI_i$ is \RTwoCcEIOwnerSD{} on source selection and \RTwoPlaceEIOwnerSD{} on front-page placement, compared with affiliate SDs of \RTwoCcEIAffSD{} and \RTwoPlaceEIAffSD. 
Owners differ as much in which sources they link to and which stories they lead with as in which events they
cover. As such, the ownership effect generalizes across multiple data sources and decision types. 
Given the possibility that model results are impacted by irregularities in the data, we also conduct a sensitivity check where we refit every model after dropping
outlet-months whose GDELT volume fell below a quarter of the outlet's 2017 monthly mean. The estimates are essentially
unchanged: the owner SD is \RTwoVolEIOwnerSD{} for $EI_i$ (against \RTwoEIOwnerSD{}), \RTwoVolEIHOwnerSD{} for $EI^H_i$
(against \RTwoEIHOwnerSD{}) and \RTwoVolIPOwnerSD{} for $IP_i$ (against \RTwoIPOwnerSD{}), and the affiliate SDs stay
small.

\begin{figure*}
    \centering
    \includegraphics[width=\textwidth]{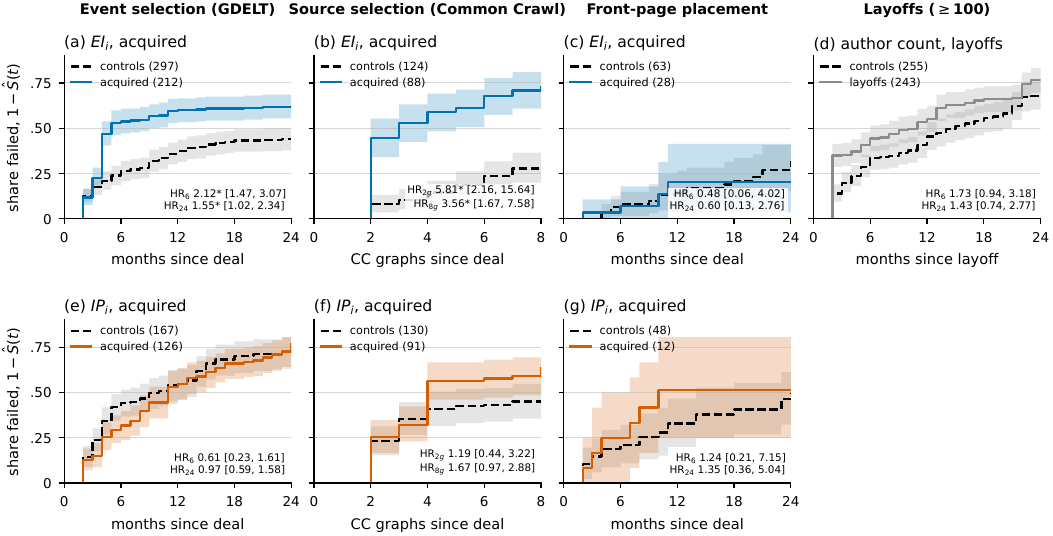}
    \caption{Survival curves for acquisitions and layoffs show the cumulative share of treatment and control groups that experience a sustained loss. Sample sizes and Cox hazard ratios at 6 and 24 months (2 and 8 graphs for Common
    Crawl) are given.}
    \label{fig:hazard_km}
\end{figure*}

\subsection{Acquisitions and organizational change}
\label{sec:acquisitions}


We follow every outlet acquired or merged into a new owner between 2018 and 2023, each matched with at least one
never-acquired outlet (within the period of observation) of the same chain tier (i.e., part of a multi-outlet group or not), same political bias group (left, center, or right), and same decile of pre-deal size (in terms of mean byline author counts over the six months before the event).

With this design we ask two questions. First, does the acquired
outlet lose independence sooner than its matched controls? Here, we define failure as a \emph{sustained} loss of at least two consecutive months where the treated outlet's measure is $>1$ SD below its pre-event level. 
We estimate Cox hazard ratios with follow-up truncated at 3, 6, 12 and 24 months post-event. 
Second, does the acquired outlet converge on its new owner? 
We track the outlet's NMI to its new sister outlets, and the share of the pre-deal gap to the acquirer's coverage profile that it closes, against targets of deals that failed. 
While similar dynamics have been identified by \citet{rappaz_dynamic_2019}, their measures are not interpretable and rates of convergence cannot be established with this method.
We further extend our matched design to large layoffs of at least 100 jobs, and add a non-editorial outcome: byline author counts.

In a time-to-event (survival) analysis, each outlet is followed from the event until its first sustained
loss or the end of the follow-up period. 
\autoref{fig:hazard_km} plots Kaplan--Meier estimates \citep{kaplan1958nonparametric} of
the cumulative share of outlets that have experienced a sustained loss by each month since the event, for treated outlets and their matched controls. Because outcomes fluctuate, controls may fail too, and so evidence for organizational influence is the gap between the treatment and control curves. A solid curve above the dashed one means treated outlets lose the outcome sooner, and the vertical gap at a month is the excess share of treated outlets that have failed by then. 
The Cox proportional-hazards model \citep{cox1972regression} summarizes the gap as a hazard ratio, the rate at which treated outlets fail relative to the matched controls, with a ratio of 2 meaning twice the controls' rate, and 0.5 half of it. 
Truncating follow-up at 3, 6, 12 and 24 months reveals change over time, as organizational influences on each of the measures may follow different time scales.
\autoref{fig:hazard_forest} shows these ratios on a log scale, one row per outcome
and horizon. As we will see, these results reinforce the findings from the Kaplan--Meier estimates in \autoref{fig:hazard_km}.

\begin{changed}
\paragraph{Acquisitions.}
Acquired outlets lose editorial independence sooner than matched never-acquired outlets (\autoref{fig:hazard_km},
\autoref{fig:hazard_forest}). Within six months \RThreeKmAcqEITreatSix\% of acquired outlets have experienced a sustained loss, against \RThreeKmAcqEICtrlSix\% of matched controls. The hazard of a sustained fall in $EI_i$ is \RThreeHzEcAcqEISix{} [\RThreeHzEcAcqEISixLo,
\RThreeHzEcAcqEISixHi] times that of the controls at six months and \RThreeHzEcAcqEITwentyFour{}
[\RThreeHzEcAcqEITwentyFourLo, \RThreeHzEcAcqEITwentyFourHi] at 24 months (\RThreeHzAcqNTreated{} acquired outlets,
\RThreeHzAcqNControl{} controls). The loss replicates on source selection, where it is larger (\RThreeHzCcAcqEITwoGraphs{}
[\RThreeHzCcAcqEITwoGraphsLo, \RThreeHzCcAcqEITwoGraphsHi] two Common Crawl graphs, about six months, after the deal;
\RThreeHzCcAcqEINTreated{} acquired outlets). Acquisitions do not appear to have a substantial impact on front-page placement, although this measure covers too few acquired outlets
(\RThreeHzPlaceAcqEINTreated) to test for significance.


Turning to media type, while all outlets are measured according to their digital footprint, much of their content may come from their medium of origin. 
However, we observe that the drop in $EI$ associated with acquisitions is consistent across outlets regardless of whether they are primarily broadcast or print outlets. 
For outlets that began as television stations, \RThreeAlphaTVSix{} of the pre-deal gap to their acquirer's coverage is closed within six months; for newspapers, \RThreeAlphaPrintTwelve{} of the same gap is closed within a year. We come back to the question of organizational influence by media type in Section~\ref{sec:validation}.

Furthermore, we can show that content convergence is a result of acquisitions, because coverage gaps do not close when deals fail. In these cases (\RThreeBetaFailedDeals{} blocked or abandoned deals), target outlets close none of the pre-deal gap to the acquirer (\RThreeBetaFailedTwelve{} after one year).
Likewise, the internal pluralism of acquired outlets does not change significantly (\RThreeHzEcAcqIPSix{}
[\RThreeHzEcAcqIPSixLo, \RThreeHzEcAcqIPSixHi] at six months).
In terms of productivity and engagement, we find that acquisitions do not negatively impact either.  In fact, acquired outlets are less likely than their controls to lose byline authors (\RThreeHzEcAcqAuthorsTwentyFour{}
[\RThreeHzEcAcqAuthorsTwentyFourLo, \RThreeHzEcAcqAuthorsTwentyFourHi] at 24 months). This may be related to pre-acquisition budget cuts leading to job losses before deals close.

\paragraph{Evaluating acquisition policy.}
\label{sec:comparing}

To ask whether regulators stop the deals that
matter editorially, we project each deal's effect on ecosystem independence over 12 months from the convergence we
measure: an approved deal is compared with a no-deal path in which its targets drift like targets of failed deals, and
a blocked deal with a with-deal path in which its targets close the measured share of the gap to the would-be acquirer.
Over \RFiveNBlockedDeals{} blocked and \RFiveNClearedDeals{} cleared deals, the projected loss is larger for blocked
deals (difference \RFiveDecisionRawDiff, $p = \RFiveDecisionRawP$ over the three independent regulatory decisions)
suggesting that regulators block large
deals rather than editorially worse ones. The approved deals are small in isolation, yet combined, their projected loss is
about 0.05 in ecosystem independence (\RFiveApprovedSumSD{} month-to-month standard deviations). 

\paragraph{Layoffs.}
Layoffs act on staff rather than on editorial independence. Outlets that cut at least 100 jobs lose byline authors
more often than their controls (\RThreeKmLayoffAuthorsTreatSix\% against \RThreeKmLayoffAuthorsCtrlSix\% by six months).
The lower hazard of $IP$ loss replicates on source selection (\RThreeHzCcLayoffIPFourGraphs{}
[\RThreeHzCcLayoffIPFourGraphsLo, \RThreeHzCcLayoffIPFourGraphsHi]), but not on front-page placement,
where the three-month ratio goes the other way (\RThreeHzPlaceLayoffIPThree{} [\RThreeHzPlaceLayoffIPThreeLo,
\RThreeHzPlaceLayoffIPThreeHi], \RThreeHzPlaceLayoffIPNTreated{} outlets, see \autoref{fig:hazard_forest_all} in the appendix).

Here, significant author count losses within the control group are indicative of the persistence of small-scale job cuts across the media industry. 
These types of job cuts are difficult to track in a dataset because organizations are not required to disclose them under the federal WARN Act.
As such, matched controls eventually lose staff at the same rate: outlets without a large announced layoff shrank too, only without the initial cut.

\begin{figure}
    \centering
    \includegraphics[width=\linewidth]{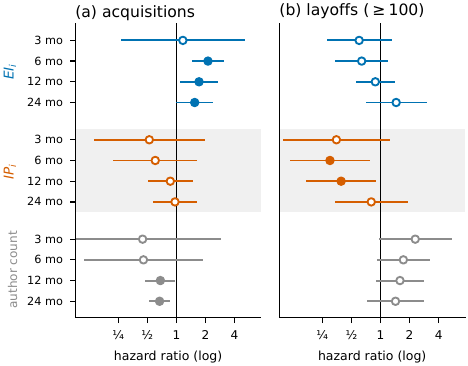}
    \caption{Cox hazard ratios [95\%] of a sustained loss after (a) acquisitions and (b) layoffs of at least 100
    jobs, with follow-up truncated at 3, 6, 12 and 24 months.
    A hazard ratio $>1$ means that treated outlets lose the outcome sooner than controls (filled markers exclude 1).}
    \label{fig:hazard_forest}
\end{figure}

\end{changed}

\subsection{Organizational influence on broadcast TV}
\label{sec:validation}
\begin{figure}
    \centering
    \includegraphics[width=\linewidth]{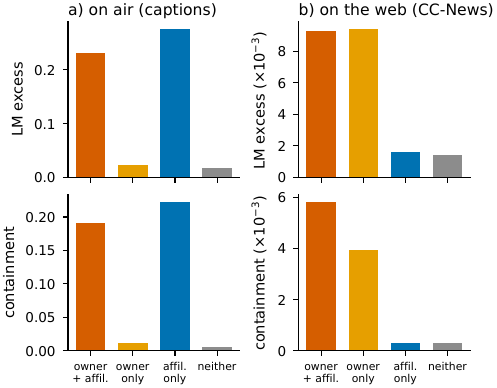}
    \caption{Owner versus affiliation redundancy on air (broadcast captions \citep{sood2022captions}) and on the same stations' websites (CC-News \citep{nagel2016ccnews}). Trends on air are in opposition to those observed on the web, with affiliation, not ownership, accounting for shared broadcast coverage (as measured by LM-surprisal).}
    \label{fig:web_vs_broadcast}
\end{figure}

Our measures observe news on the Web, while television groups also coordinate on air. Comparing the same station pairs on air (closed captions) and on their websites
(\autoref{fig:web_vs_broadcast}) shows the two layers are organized differently.
Here, closed captions from broadcast TV come from the Internet Archive's TV News Archive, spanning 2015--2016 \citep{internetarchive_tvnews,sood2022captions}, and news articles scraped from news websites come from the Common Crawl News dataset (CC-News), spanning 2016--2024
\citep{nagel2016ccnews}. Analogous to the pairwise NMI measure, we calculate pairwise linguistic redundancy between two outlets as the drop in a language model's per-token surprisal of one station's text when the other's same-day text is in context. 

First, on the web, the surprisal measure corroborates the pairwise NMI findings for the ownership effect on digital news (as in \autoref{fig:r1_similarity_by_coorg}), where
ownership adds redundancy within a network (\RFourWebExcessOwnerWithinNet, $p<.001$) and affiliation adds almost none. 
On air, however, redundancy follows the network affiliation, where co-affiliated outlets share \RFourAirExcessNetAmongDiffOwner{} more surprisal reduction than random outlets
and co-owned outlets share almost nothing (\RFourAirExcessOwnerWithinNet). 
The specific ownership effects we report are therefore a property of digital media. For other media types, affiliation and content syndication may play a more important role.



\subsection{Longitudinal analysis}


We find that platform decisions, not ownership, seem to dominate the ecosystem-level engagement and newsroom employment measures (\autoref{fig:totals_timeseries}).
Organic search traffic rises from late 2020 as outlets turn to search, then falls after
Google's US launch of AI Overviews, which reduce clicks to linked sites \citep{chapekis2025aisummaries}, from an estimated 2.8 billion monthly visits in May 2024 to 1.6 billion in June 2026. Summed over outlets, byline authors decline throughout, from 157{,}000 in March 2018 to 38{,}000 in September 2026. Our byline author count trendline correlates highly with monthly US newspaper employment data ($r=0.867$) compiled by the Bureau of Labor Statistics \cite{bls2026newspapers}.
In contrast, and despite the pronounced changes that we describe at the individual outlet level, our ecosystem-level editorial measures of external pluralism ($EP$) and $EI$ are stable over the period. EP varies little (monthly SD $\le$ 0.03 for every decision), and the information-weighted mean of $EI_i$ is steady throughout, with an SD of 0.05 for event selection.

\section{Discussion}

Our results answer the four questions we set out with. First, ownership influences editorial decisions on the Web.
Co-owned outlets share \ROneGdeltPooledSameOwnerRatio{} times the information of the shuffled-owner null in event
selection, \ROneCcPooledSameOwnerRatio{} times in source selection and \ROnePlacePPooledSameOwnerRatio{} times in
front-page placement (\autoref{fig:r1_similarity_by_coorg}). Second, ownership matters far more than network affiliation: the owner SD of $EI_i$ is
\RTwoEIOwnerSD{} against \RTwoEIAffSD{} for affiliates (\autoref{fig:gamm_coefficients_for_ownership}). Only on broadcast
television does affiliation dominate (\autoref{fig:web_vs_broadcast}). Third, acquisitions reduce editorial independence while layoffs do not. Acquired
outlets lose $EI_i$ sooner than matched controls (\autoref{fig:hazard_km}) and close most of
the coverage gap to their acquirer within a year, while outlets involved in \RThreeBetaFailedDeals{} failed deals do not move. Layoffs cut staff
instead (\autoref{fig:hazard_forest}). Fourth, our projections suggest that regulators tend towards blocking large deals rather than editorially harmful ones: per target outlet, our projections suggest that blocked deals would not have been more determinant to independence than approved deals, although this rests on few regulatory decisions and on projections rather than observed outcomes.

Two deals in particular show that regulatory considerations can and should extend beyond size and reach. In 2018 the FCC
stopped Sinclair Broadcast Group's \$3.9 billion acquisition of Tribune Media, which would have taken an owner of
over 190 local stations \citep{sinclair} to about 70\% of US households; its review did not measure editorial
independence, although that April a compilation of Sinclair anchors reading the same scripted message had gone viral
\citep{sinclair_backlash} and Sinclair's outlets are the least independent of the \RTwoEIOwners{} owners in our model
(owner random effect \RTwoEILowestOwnerBLUP{} outcome SDs). The GateHouse--Gannett merger of 2019, a print deal outside
the FCC's remit, was cleared without any editorial review. The impact of each cleared deal like this is small (about 0.001 in ecosystem
independence), but together our projections put their cost at just over 1 SD of our EI measure.

These findings extend a literature built on single owners and single decisions \citep{theine2025does}.
\citet{martin2019local} found that Sinclair-acquired stations shifted their coverage toward national politics. We
observe ecosystem-wide convergence on the acquirer across three different editorial decisions, and further show that outlets involved in failed deals are not impacted.
\citet{gentzkow2010drives} found that newspaper slant adapts to readers, not owners. Our measures show that when it comes to coverage decisions, ownership matters. 
While post-acquisition convergence in news coverage had already been demonstrated by \citet{rappaz_dynamic_2019} for a sample of $n=12$ outlets, their measures are not interpretable, so rates of convergence cannot be established. 
We extend the idea behind their case study to every acquisition in the panel, with rates that can be compared across owners and decisions.

Our findings have several implications. 
First, editorial independence measures require formal information-theoretic definitions of redundancy. When multiple Sinclair anchors read from the same script
\citep{sinclair_backlash}, the combination of editorial control with a lack of transparency enables audience manipulation. Measures that
count duplicated content, such as recycling \citep{hendrickx2021why} or our $EI^H_i$, cannot separate these instances from benign AP news wire usage. Wire copy lowers $EI^H_i$  (our definition of a content recycling measure) by about three times as much as $EI_i$, and so content duplication measures reads AP wire use as dependence on the owner.
What's more, after an acquisition $EI_i$ falls while $EI^H_i$ does not, so the loss lies in choices specific to the new owner, which a duplication measure misses.
Separating owner-specific from shared-agenda information is what lets $EI_i$ track independence.

Second, organizational influence depends on the medium. On air, stations’ shared content follows their affiliation. On their websites, it follows their owner (Section~\ref{sec:validation}). While broadcast regulation counts stations and audience reach, treating network affiliation as the content content syndication mechanism, this is not the case on the web. Here, the owner is the stronger influence. A board for CBS News and CNN would therefore need to watch both the affiliate’s feed on air and the owner’s choices online.

Third, the window for such oversight is short and harms from editorial negligence accumulate. 
Acquired outlets the coverage gap
to their new owner within a year, so remedies applied after a deal closes
arrive once editorial changes have already begun. And because each cleared deal is small on its own while together they cost about
0.05 SDs of ecosystem-level independence, reviewing each deal separately hides the ecosystem effect. Editorial measures computed
continuously from Web-scale data allow both the screening of individual deals before they close,
and the tracking of ecosystem-level effects across multiple concurrent mergers and acquisitions.
Of course, our measures are not without limitations. First, our evidence is mostly observational and not causal. If co-owned outlets decide alike, this does
not imply that owners made them do so. In addition, chains may have a preference towards acquisitions that already align with their outlets, or whose audience is perhaps more suited towards the same viewpoints. Although existing FCC regulations guard against a single company owning multiple television stations in the same locale (often called the duopoly rule), if co-owned outlets do share regions and audiences, our measure may mistake their locality for dependence. A locale term would need to be added to our decomposition to prevent this (\ref{eq:pid}).
In general, reading PID and PED atoms as independence and pluralism is a modeling choice (Appendix~\ref{app:mi}), and many other valid definitions may exist that also address the same issues of separating group level redundancy from ecosystem level redundancy.

There are also several data issues that impacted our study. 
The filter removes events covered by fewer than three outlets in order to address overly granular GDELT event codings may also filter out articles with local stories or investigative and niche reporting.
What's more, GDELT stopped crawling many chain-owned sites after ownership changes resulted in their crawlers being blocked, and we are forced to mask these gaps.
Front-page placement is particularly affected, having coverage over fewer outlets than GDELT events and the CommonCrawl webgraph offer. 
In addition, starting in October 2025 the front-page data pipeline no longer appears to be actively maintained by GDELT, limiting its future use.


Several extensions follow. The FCC already weighs viewpoint diversity, but through ownership counts and
reach; our projections suggest screening mergers on editorial measures as well, and more research should be done to support this. Given that the information-theoretic measures generalize to groupings of any kind, the influence that of chain tier, private-equity ownership, or political alignment has over editorial decisions can be studied using this framework. 
Finally, given the international scope of GDELT and Common Crawl, the analysis done here may be extended to other countries. To support these efforts, we release the measures, the code and the owner-at-month panel on publication.

\section{Conclusion}
Owners influence what news sites cover, cite, and feature, and acquisitions move outlets toward their new owner
within months. Through web-scale datasets and our information-theoretic framework, editorial independence can now be measured.

\section{Acknowledgements}

We thank Dr. Gregory Martin and for providing an acquisition dataset and Dr. Zachary Metzger at the Medill Institute for providing ownership data for print outlets.
This work was supported in part by the Office of Naval Research grant (N000142412568) and the Knight Foundation. 
The views and conclusions in this document are those of the authors and should not be interpreted as representing the official policies, either expressed or implied, of the Knight Foundation, Office of Naval Research, or the U.S. Government.

\bibliographystyle{plainnat}
\bibliography{Tradeoff,other,div_paper,methods_dvs,related_work}

\appendix

\begin{figure*}
    \centering
    \includegraphics[width=\textwidth]{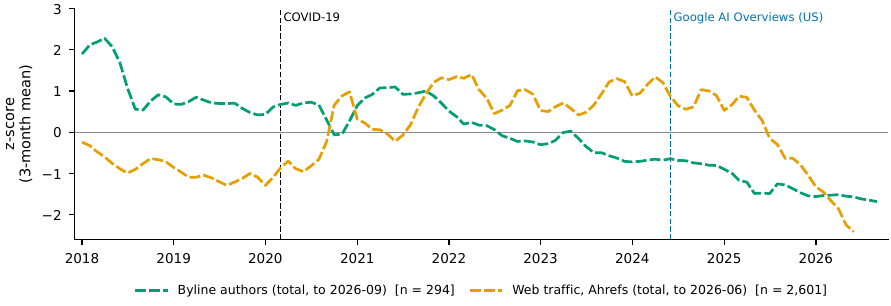}
    \caption{Productivity and engagement over time: total byline authors (294 outlets, to 2026-09), Ahrefs
    estimates of organic web traffic (2,601 outlets, to 2026-06), each summed
    over a fixed outlet set, z-scored over its own window and smoothed with a centered 3-month mean.
    Markers: start of COVID-19 lockdowns in the US (March 2020) and Google's US launch of AI Overviews (May 2024).}
    \label{fig:totals_timeseries}
\end{figure*}

\section{News ownership labels}
\label{app:ownership}

\paragraph{Construction.}
Owner labels come from crowd-sourced lists of media-company assets on Wikipedia. For each broadcasting or publishing
company with a Wikipedia category of the outlets it owns (for example, the category of Nexstar Media Group), we
retrieve the member pages with the Petscan service. Each outlet's official domain is taken from Wikidata through a SPARQL
query, with the website field of the Wikipedia infobox as a fallback. Domains are resolved through HTTP redirects to
their canonical form and normalized (lower case, without \texttt{www.} or paths). Outlets are mostly newspapers,
television stations and radio stations. Network-affiliate labels (for example, ABC or NBC affiliation of a local television
station) are built the same way, from Wikipedia's network-affiliate categories; an outlet can carry several. The same
procedure, applied worldwide and filtered with web-traffic estimates and site categories, produces a multi-country
news-ownership dataset; this paper uses only its US news outlets, plus the affiliate labels, which that dataset lacks.

\paragraph{Validation on US outlets.}
The labels were validated against human annotation and against an independent ownership directory. On a sample of
321 US news domains annotated by hand with their owner, the Wikidata owner labels agree with the annotation for
95.8\%. The Medill Local News Initiative directory of US newspapers \citep{LocalNews2024,LocalNews2025} provides a
second check, on the 638 domains where its print titles coincide with our news websites. Our owner labels agree with
Medill's for 73.5\% of these domains, which receive 91.9\% of the overlap's web traffic. Where the two disagree, Medill
is usually right, because Wikipedia lags recent acquisitions (for example, the sale of Gannett titles to CherryRoad
Media \citep{GulbanCherryRoad2025}). Counting either Medill's current or its previous owner as a match raises
agreement to 93.6\% of domains and 98.8\% of traffic. The labels are therefore accurate up to the timing of recent
sales, which the owner-at-month correction below addresses. These figures are for the US outlets of the multi-country
collection, built like our labels.

\paragraph{Owner at month.}
The Wikipedia labels are a snapshot compiled in 2023--2025 and carry no dates, so they miss the deals made during our
window (among them GateHouse--Gannett, Nexstar--Tribune, Gray--Raycom and Lee--BH Media). For every one of the 1,364
owner-labeled outlets we therefore researched the owner on 2015-01-01 and on 2026-09-30 and every change in between,
with its closing date and a source. The sources are Wikipedia station and newspaper histories (1,788 of 2,318 owner
spells), SEC filings and company press releases (165), and trade press, the Medill directory snapshots for 2014, 2016,
2018, 2020 and 2025, and live site checks (365). Every outlet's starting owner was checked against at least one source,
and 1,755 of the 2,318 spells are rated high confidence. The result is a panel of 2,318 owner spells over 278 dated deals.
It corrects 99 snapshot labels (51 outlets never held by the labeled owner in the window, 48 sold or closed before the
snapshot), and 667 outlets change owner at least once in 2015--2026. As a cross-check, 123 of the 135 in-window
acquisitions recorded for these outlets in FCC ownership-transfer records coincide with a
change in the panel within two months. The other 12 are explained by license versus operating-control timing, sales of
non-license assets, closings before the window, month offsets, or sub-brands.

\paragraph{Conventions.}
The owner of month $M$ is the owner at the end of the first day of $M$, so a deal switches an outlet's owner, and its
own group $G_{c(i)}$, from the first month that starts on or after the closing. Stations run under local marketing or
shared-services agreements carry the operating group rather than the licensee. Events at the level of a parent company
that do not change the operating owner (for example, the CBS--Viacom merger) leave the label unchanged. Predecessor owners
keep their own names, so a merger splits an owner's history into a pre-deal and a post-deal owner; this is why the
central model has more owner levels than there are owners today (Section~\ref{sec:model-gamm}). Closed outlets have no owner after closure.

\section{Data and decision details}
\label{app:data-details}
\paragraph{Front-page placement.} The Global Frontpage Graph records each homepage link's position as a percentage of
the page's links, from 0 (first in the page) to 100 (last). For one Sunday a month (March 2018 to December 2024) we
match each outlet's homepage links to the article URLs GDELT links to events in the three days up to that Sunday, and
keep each event's best (highest) position on the page. An outlet is active if at least 50 events appear on its front
page that Sunday; $r_i(e)$ is the event's rank among them and $n_i$ their number. The outlet measures use front-page
presence, $X_i(e)=1$ if $e$ is on $i$'s front page, because the PID requires a binary coverage variable. Pairwise
comparisons use four levels, $P_i(e)=0$ absent, $1$ lower page ($r_i(e)\ge n_i/2$), $2$ upper page, $3$ lead (the top
$k=\ROnePlacePLeadK$ positions), with NMI computed over events on at least three front pages.
\paragraph{Organizational events.} \emph{Layoffs}: 451 announced job cuts at US news organizations, compiled by hand
from public trackers of newsroom cuts and coded from the report or announcement cited for each entry (outlet, date,
jobs cut, closure); we use cuts of at least 100 jobs. \emph{Acquisitions}: changes of owner in the owner-at-month panel
(Appendix~\ref{app:ownership}), which draws on the television-station ownership records of \citet{martin2026media} and
the Medill newspaper ownership records \citep{LocalNews2025}; we use acquisitions and mergers from 2018 to mid-2023.

\section{Measure and model details}
\label{app:methods-details}

\subsection{The shared agenda}
The estimand is how outlets divide attention among items that receive attention from several of them, not pluralism over all news. Two consequences are handled explicitly. First, without the filter the joint distribution over thousands of binary variables is so sparse that almost every coverage pattern is unique and plug-in entropy estimates count items rather than structure. Second, conditioning on the column sum in \eqref{eq:filter} induces positive dependence among the $X_i$ even when outlets choose independently; every measure is therefore reported against a null model that preserves the filter (Section~\ref{sec:nulls}).

The field source is a count rather than an indicator because, under the filter \eqref{eq:filter}, the event that some outside outlet covered $e$ is almost always true and would carry no information; the count records the item's prominence outside $i$'s group.

\subsection{Partial information decomposition}
We use the two-source measure of \citet{bertschinger2014quantifying} (BROJA), for which unique information is defined by a convex optimization over distributions sharing the observed source--target marginals; with two sources the leading PID measures agree on the qualitative behavior of these atoms, and the aggregation of the own group to a single binned count in \eqref{eq:sources} is what keeps the decomposition at two sources. The four atoms together with the unexplained remainder $H(X_i \mid Z^{\own}_i, Z^{\fld}_i)$, the coverage predictable from neither source, partition $H(X_i)$. The four explained atoms divided by $I_i$ form outlet $i$'s \emph{compositional profile} under $C$; they sum to one, and the explained fraction $I_i / H(X_i)$ is retained so that any atom can be re-expressed as a share of $H(X_i)$.

\subsection{Estimation}
\label{sec:estimation}

Marginal entropies and the joint tables of $(X_i, Z^{\own}_i, Z^{\fld}_i)$ are estimated by plug-in over the items
in $E_t^{(m)}$. The outlet-level PID uses the BROJA measure, computed with our own solver of its convex program and
validated against the \texttt{dit} implementation to $2\times10^{-3}$ bits; pairwise mutual information is plug-in from
$2\times 2$ tables. Plug-in dependence carries a positive bias of order $(\text{cells}-1)/(2n \ln 2)$, largest for
outlets with small $H(X_i)$, which the null models below calibrate.

\subsection{Null models and reporting}
\label{sec:nulls}

Two null models calibrate the measures. A degree-preserving randomization of the outlet-by-item incidence on
$E_t^{(m)}$ (Curveball \citep{strona2014fast}; 200 draws for the pairwise measures, 30 for the PID measures) keeps each
outlet's coverage volume and each item's coverage count, so the filter \eqref{eq:filter} and the dependence it induces
are present in the null. A permutation of the group labels $c(i)$ with the incidence fixed turns a within-group statistic
into evidence about the grouping; the pairwise comparisons of \autoref{fig:r1_similarity_by_coorg} use it. The models
and hazard analyses use the raw series.

\subsection{Central model}
In \eqref{eq:gamm}, $\mathrm{ap}_{it}$ indicates that the outlet-month published at least one authored article and
$\mathrm{share}_{it}$ is the share of those articles carrying an Associated Press byline (read as
zero when no wire-share record exists), so that $\beta_2$ is the wire-dependence slope among months with authored
output; the information-theoretic outcomes carry no political-bias term, since the bias label enters
their construction; $f(t)$ is a thin-plate smooth of calendar time with basis
dimension 12; $u_{o(i)} \sim \mathcal{N}(0, \sigma_o^2)$ is a random intercept for the parent owner $o(i)$, the owner at month $t$ (Appendix~\ref{app:ownership});
$a_{n(i)} \sim \mathcal{N}(0, \sigma_a^2)$ is a random intercept for the network affiliate, present only for
affiliated outlets; and $g_i(t)$ is a per-outlet factor smooth of time with basis dimension 5, which absorbs each
outlet's own level and trend so that the ownership terms are identified from what co-owned outlets share beyond
their individual histories. Counts enter as $\log(1 + y)$, engagement as $\operatorname{asinh}(y)$, and shares,
distances and entropies on the identity scale. Outlet-months in GDELT coverage
gaps are dropped.

\paragraph{Serial dependence.} Outlet-months are not independent draws. After the full fit we compute each
outlet's lag-one residual autocorrelation; when the median across outlets exceeds $0.30$ the model is refitted
with a first-order autoregressive error at the mean residual autocorrelation, and every comparison below is made
between models with the same error structure.

\paragraph{Variance share.} The variance share of a term is the variance of its fitted contribution as a share of
the variance of the full fitted predictor plus the residual variance; it is an effect size, not evidence. Slopes are
reported in standard deviations of the outcome.

\section{Shared information with groups of outlets}
\label{app:group-size}
\begin{figure}
    \centering
    \includegraphics[width=\linewidth]{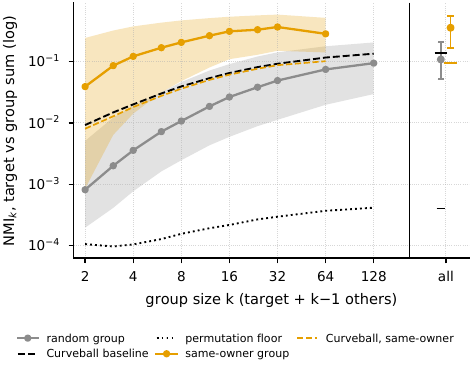}
    \caption{Event selection (GDELT): NMI between an outlet's coverage and the binned summed coverage of $k-1$ other
    outlets, by group size $k$; ``all'' uses every other outlet (for same-owner groups, all co-owned outlets). Lines are
    medians and bands interquartile ranges over 300 draws per $k$ in each of the 84 monthly windows. Dashed: the same
    draws on Curveball (degree-preserving) randomizations. Dotted: the plug-in bias floor ($X_i$ shuffled over events).
    Same-owner groups are shown where at least three owners have $k$ outlets.}
    \label{fig:nmi_vs_group_size_gdelt_events}
\end{figure}

Pairs of unrelated outlets share little event selection (\autoref{fig:r1_similarity_by_coorg}a). To test whether
groups share more, we hold out an outlet $i$, draw $k-1$ other outlets, and compute the NMI between $X_i$ and the
binned count of those outlets covering each event, built like the field source; $k=2$ is the pairwise
NMI of Eq.~\ref{eq:nmi}. Shared information rises steeply with group size: from a median of 0.0008 for one random outlet to 0.011
for seven and 0.11 for all other outlets (\autoref{fig:nmi_vs_group_size_gdelt_events}). This rise reflects event
popularity, not coordination. On a degree-preserving randomization of the outlet--event matrix (Curveball), which keeps
how many events each outlet covers and how many outlets cover each event, the curve is higher at every $k$ (0.009 for
pairs, 0.14 for the field): unrelated outlets are more independent than their coverage totals imply, as expected if local
outlets cover different markets. Groups of co-owned outlets sit four to seven times above that baseline for $k=3$ to
32; eight co-owned outlets explain about 21\% of an outlet's coverage entropy, eight random outlets about 1\%.

\section{Acquisitions and layoffs on source selection and front-page placement}
\label{app:other-events}
\begin{figure}
    \centering
    \includegraphics[width=\linewidth]{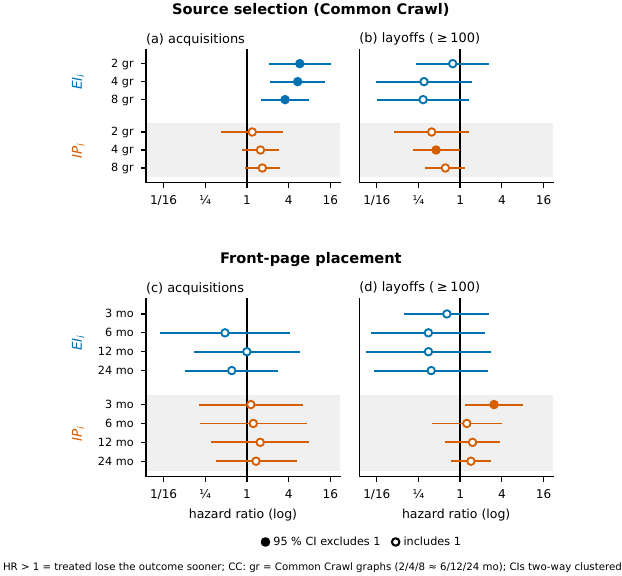}
    \caption{Cox hazard ratios [95\%] of a sustained loss of $EI_i$ and $IP_i$ measured on (a, b) source selection
    and (c, d) front-page placement, after acquisitions (a, c) and layoffs of at least 100 jobs (b, d), against matched
    controls; follow-up truncated at 2, 4 and 8 Common Crawl graphs (source selection) or 3, 6, 12 and 24 months
    (placement); intervals clustered by outlet and by event; filled markers exclude 1. Analogous to
    \autoref{fig:hazard_forest}.}
    \label{fig:hazard_forest_all}
\end{figure}

\autoref{fig:hazard_forest_all} repeats the hazard analysis of \autoref{fig:hazard_forest} with $EI_i$ and $IP_i$
measured on the other two editorial decisions; the design, matching and failure rule are unchanged, and intervals are
clustered by outlet and by event. On source selection, time is counted in Common Crawl graphs (two, four and eight
graphs, about 6, 12 and 24 months). Acquired outlets lose editorial independence in their linking sooner than their
controls (\RThreeHzCcAcqEITwoGraphs{} [\RThreeHzCcAcqEITwoGraphsLo, \RThreeHzCcAcqEITwoGraphsHi] at two graphs,
\RThreeHzCcAcqEIEightGraphs{} [\RThreeHzCcAcqEIEightGraphsLo, \RThreeHzCcAcqEIEightGraphsHi] at eight), while their
internal pluralism does not change significantly; outlets with large layoffs are less likely to lose internal pluralism
(\RThreeHzCcLayoffIPFourGraphs{} [\RThreeHzCcLayoffIPFourGraphsLo, \RThreeHzCcLayoffIPFourGraphsHi] at four graphs). On
front-page placement the samples are small (\RThreeHzPlaceAcqEINTreated{} acquired outlets for $EI_i$,
\RThreeHzPlaceAcqIPNTreated{} for $IP_i$) and every interval includes 1, except a higher hazard of losing $IP_i$ three
months after a large layoff (\RThreeHzPlaceLayoffIPThree{} [\RThreeHzPlaceLayoffIPThreeLo,
\RThreeHzPlaceLayoffIPThreeHi]).

\label{app:other-decisions}
\begin{figure}
    \centering
    \IfFileExists{fig/www/r2a_owner_effect_size_other_decisions.pdf}{%
    \includegraphics[width=\linewidth]{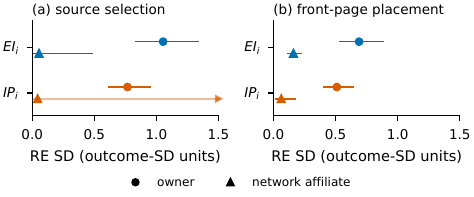}}{%
    \fbox{\parbox{0.95\linewidth}{\centering\vspace{2cm}\textcolor{blue}{r2a\_owner\_effect\_size\_other\_decisions.pdf (task D67)}\vspace{2cm}}}}
    \caption{Owner ($\bullet$) and network-affiliate ($\blacktriangle$) random-effect SDs in outcome-SD units, 95\%
    intervals, from the central GAMM refitted on (a) source selection (one window per Common Crawl graph;
    \RTwoCcEIOutlets{} outlets) and (b) front-page placement (monthly Sundays; \RTwoPlaceEIOutlets{} outlets); analogous to
    \autoref{fig:gamm_coefficients_for_ownership}a.}
    \label{fig:gamm_other_decisions}
\end{figure}


\section{Mutual-information forms of the measures}
\label{app:mi}

Each measure in Section~\ref{sec:methods} has a form written in conditional entropies, mutual information, co-information or total correlation rather than decomposition atoms. This appendix gives those forms, the exact identities that relate them to the atoms where such identities exist, and the differences where none does. The MI forms are computed on the same source variables and the same items as the decomposition forms, so the two can be compared on identical inputs. Throughout, $I_{ij} = I(X_i; X_j)$ denotes the pairwise mutual information between outlets and $s_i = \sum_{j} I_{ij} / \sum_{i,j} I_{ij}$ outlet $i$'s share of total pairwise coupling.

\subsection{Identities}
\label{app:identities}

Two identities link the atoms of \eqref{eq:pid} to standard quantities:
\begin{align}
  I(X_i ; Z^{\own}_i) &= \Unq_i^{\own} + \Red_i, \label{eq:id1}\\
  \CoI_i &\;\equiv\; I(X_i ; Z^{\own}_i) + I(X_i ; Z^{\fld}_i) - I_i \notag\\
         &= \Red_i - \Syn_i, \label{eq:id2}
\end{align}
where $\CoI$ is co-information. Identity \eqref{eq:id2} states the sense in which the MI forms are less expressive than the decomposition forms: they report redundancy net of synergy.
\subsection{Family 1}
\label{app:family1}

The MI form of internal pluralism, on the same source variable so that the relation is exact, is
\begin{equation}
  \IP^{H}_i = 1 - \frac{I(X_i ; Z^{\own}_i)}{I_i}, \quad
  \IP_i - \IP^{H}_i = \frac{\Red_i}{I_i}
  \label{eq:iph}
\end{equation}
by \eqref{eq:id1}; likewise for $\EI_i$.

\end{document}